\documentclass[onecolumn,showpacs,preprintnumbers,amsmath,amssymb]{revtex4}
\usepackage{graphicx}
\begin{document}

\title{Stochastic Turing patterns in the Brusselator model}

\author{Tommaso Biancalani}
\affiliation{Dipartimento di Fisica,  Universit\`{a} degli Studi di
Firenze, via G. Sansone 1, 50019 Sesto Fiorentino, Florence,  Italy
}
\author{Duccio Fanelli}
\affiliation{Dipartimento di Energetica, Universit\`{a} degli Studi di Firenze,
via S. Marta 3, 50139 Florence, Italy and INFN, Sezione di Firenze.
}
\author{Francesca Di Patti}
\affiliation{ Dipartimento di Fisica ``Galileo Galilei'',
Universit\`{a} degli Studi di Padova, via F. Marzolo 8, 35131
Padova, Italy}

\begin{abstract}
A stochastic version of the Brusselator model is proposed and
studied via the system size expansion. The mean-field equations are
derived and shown to yield to organized Turing patterns within a
specific parameters region. When determining the Turing condition
for instability, we pay particular attention to the role of cross
diffusive terms, often neglected in the heuristic derivation of
reaction diffusion schemes. Stochastic fluctuations are shown to
give rise to spatially ordered solutions, sharing the same
quantitative characteristic of the mean-field based Turing scenario,
in term of excited wavelengths. Interestingly, the region of
parameter yielding to the stochastic self-organization is wider than
that determined via the conventional Turing approach, suggesting
that the condition for spatial order to appear can be less stringent
than customarily believed.
\end{abstract}


\maketitle

\vspace{0.8cm}

\section{Introduction}
\label{intro}

Turing instability constitutes a universal paradigm for the
spontaneous generation of spatially organized patterns
\cite{Turing}. It formally applies to a wide category of phenomena,
which can be modeled via the so called reaction diffusion schemes.
These are mathematical models that describe the coupled evolution of
spatially distributed species, as driven by microscopic reactions
and freely diffusing in the embedding medium. Diffusion can
potentially seed the instability by perturbing the mean-field
homogeneous state, through an activator-inhibitor mechanism, and so
yielding to the emergence of  patched, spatially inhomogeneous,
density distribution \cite{buceta}. The realm of application of the
Turing ideas encompasses different fields, ranging from chemistry to
biology, from ecology to physics. The most astonishing examples, as
already evidenced in Turing original paper, are perhaps encountered
in the context of morphogenesis, the branch of embryology devoted to
investigating the development of patterns and forms in biology
\cite{murray} .

Beyond the qualitative agreement, one difficulty in establishing a
quantitative link between theory and empirical observations has to
do with the strict conditions for which the organized Turing
patterns are predicted to occur. In particular, and with reference
to simple predator-prey competing populations, the relative degree
of diffusivity of the interacting species has to be large according
to the theory prescriptions, and at variance with the direct
experimental evidence. Moreover, patterns formation appears to be
rather robust in nature, as opposed to the Turing predictive
scenario, where a fine tuning of the parameters is often necessary.

In \cite{McKane-PredPrey} it was demonstrated that collective
temporal oscillations can spontaneously emerge in a model of
population dynamics, as due to a resonance mechanism that amplifies
the unavoidable intrinsic noise, originating from the discreteness
of the system. Later on an extension of the model was proposed
\cite{LugoMcKane} so to explicitly account for the notion of space.
It was in particulary shown that the number density of the
interacting species oscillates both in time and space, a macroscopic
effect resulting from the amplification of the stochastic
fluctuations about the time independent solution of the
deterministic equations. More recently, Butler and Goldenfeld
\cite{Goldenfeld} proved that persistent spatial patterns and
temporal oscillations induced by demographic noise, can develop in a
simple predator-prey model of plankton-herbivore dynamics. The model
considered by the authors of \cite{Goldenfeld} exhibits a Turing
order in the mean field theory. The effect of intrinsic noise
translates however into an enlargement of the parameter region
yielding to the Turing mechanism, when compared to its homologous
domain predicted within the conventional linear stability analysis.
This is an individual based effect, which has to be accommodated for
in any sensible model of natural phenomena, and which lacks in the
Turing interpretative scenario, that formally applies to the
idealized continuum limit. Interestingly, as reported in
\cite{DeAnna}, the discreteness of the scrutinized medium can yield
to robust spatio-temporal structures, also when the system does not
undergo Turing order in its mean-field, deterministic version.

Starting from this setting, we present  the results of our
investigations carried out for a spatial version of the Brusselator
model, which we shall be introducing in the forthcoming section. As
opposed to the analysis in \cite{Goldenfeld}, we will operate within
the so called urn protocol, where individual elements belonging
to the inspected species are assumed to populate an assigned
container. The proposed formulation of the Brusselator differs from
the one customarily reported in the literature. Additional terms
are in fact obtained building on the underlying
microscopic picture. These latter contributions are generally
omitted, an assumption that we interpret as working in a diluted
limit. Interestingly, the phase diagram of the homogeneous
(aspatial) version of the model, is remarkably different from its
diluted analogue, a fact we will substantiate in the following.

Furthermore, when constructing the mean-field dynamics from the
assigned microscopic rules, one recovers non-trivial cross diffusive
terms, as previously  remarked in \cite{LugoMcKane}. These are
derived within a self-consistent analysis and ultimately stands from
assuming a finite carrying capacity in a given spatial patch. As
such, they potentially bear an important physical meaning which
deserves to be further elucidated. This scheme, alternative to
conventional reaction-diffusion models, calls for an extension of
the original Turing condition, that we will derive in the following.
This is the second result of the paper.

Finally, by making use of the van Kampen  system size expansion, we
are able to estimate the power spectrum of fluctuations analytically
and so delineate the boundaries of the spatially ordered domain in
the relevant parameters' space. As already remarked in
\cite{Goldenfeld}, the role played by the intimate grainess can
impact dramatically the Turing vision, returning a generalized
scenario that holds promise to bridge the gap with observations.

\section{A spatial version of the Brusselator model}
\label{model}

The Brusselator model represents a paradigmatic example of an
autocatalytic chemical reaction. This scheme was originally devised
in 1971 by Prigogine and Glandsdorff \cite{PG} and quickly gained
its reputation as the prototype model for oscillating chemical
reaction of the Belosouv-Zhabotinsky \cite{BZ,strogatz} type. In the
following we shall consider a slightly modified version of the
original formulation, where the number of reactants $X$ and $Y$ is
conserved and totals in $N$, including the empties, here called $E$
\cite{McKane-PredPrey}. Moreover we will consider a spatially
extended system composed of $\Omega$ cells, each of size $l$, where
reactions are supposed to occur \footnote{Working with the empties
$E$ implies imposing a finite carrying capacity in each micro-cell.
This procedure defines the so called ``urn protocol'', and allows
for a technically easier implementation of the van Kampen expansion
as presented in \cite{McKane-PredPrey}. On the other hand, and
besides technical reasons, assuming a limited capacity in space is a
reasonable physical request. It is therefore interesting to explore
how such a choice reflects on the subsequent analysis. Let us
anticipate that the presence of the empties $E$ will sensibly modify
the  mean-field phase diagram of the Brusselator model, as concerns
both the homogeneous and the spatial dynamics. Conversely, the
observation that the Turing region gets enlarged by demographic
noise holds true, irrespectively of the specific urn representation
here invoked \cite{Goldenfeld}.}. Practically each cell hosts a
replica of the Brusselator system: The molecules are however allowed
to migrate between adjacent cells, which in turn implies an
effective spatial coupling imputed to the microscopic molecular
diffusion. A periodic geometry is also assumed so to restore the
translational invariance. Although the calculation can be carried
out in {\it any} space dimension $D$ (see \cite{LugoMcKane, DeAnna})
we shall here mainly refer to the $D=1$ case study, so to privilege
the clarity of the message over technical complications. It should
be however remarked that our conclusions are general and remain
unchanged in extended spatial settings. Mathematically the model can
be cast in the form:
\begin{eqnarray*}
A + E_i &\overset{a}{\rightarrow}& A + X_i, \\
X_i + B &\overset{b}{\rightarrow}& Y_i + B, \\
2X_i + Y_i &\overset{c}{\rightarrow}& 3X_i, \\
X_i &\overset{d}{\rightarrow}& E_i,
\end{eqnarray*}
where the index $i$ runs from $1$ to $\Omega$ and  identifies the
cell where the molecules are located.  The autocatalytic species of
interest to us are $X_i$ and $Y_i$. The elements $A$ and $B$ work as
enzymatic activators, and keep constant in number. As such, they can
be straightforwardly absorbed into the definition of the reaction
rates. Let us label with $n_i$ (resp. $m_i$ and $o_i$) the number of
element of type $X_i$  (resp. $Y_i$ and $E_i$) populating cell $i$.
Then, assuming $N$ to identify the maximum number of available cases
within each cell, one gets $N=n_i+m_i+o_i$ a condition which can be
exploited to reduce the actual number of dynamical variables to two.
The dynamics of the model is ultimately related to studying the
coupled interaction between the discrete species $n_i$ and $m_i$.
The migration between neighbors cells is specified through:
\begin{eqnarray*}
X_i + E_j &\overset{\mu}{\rightarrow}& E_i + X_j, \\
Y_i + E_j &\overset{\delta}{\rightarrow}& E_i + Y_j.
\end{eqnarray*}
Assuming a perfect mixing in each individual cell $i$, the
transition probabilities $T(\cdot | \cdot)$ read
\cite{McKane-PredPrey}:
\begin{equation}
\begin{split}
\label{spacetransprob}
T(n_i+1,m_i|n_i,m_i) &= a \; \frac{N-n_i-m_i}{N \Omega}, \\
T(n_i-1,m_i+1|n_i,m_i) &= b \; \frac{n_i}{N \Omega}, \\
T(n_i+1, m_i-1|n_i,m_i) &=   d\frac{{n_i}^2m_i}{N^3 \Omega}, \\
T(n_i-1, m_i|n_i,m_i) &=  c \; \frac{n_i}{N \Omega},
\end{split}
\end{equation}
where according to the standard convention,  the rightmost input
specifies the original state and the other entry stems for the final
one. In addition, the migration mechanism between neighbors cell is
controlled by:
\begin{equation}
\begin{split}
\label{migrprob}
T(n_i-1,n_j+1|n_i,n_j) &= \mu \; \frac{n_i}{N}\frac{N-n_j-m_j}{N \Omega z}, \\
T(m_i-1,m_j+1|m_i,m_j) &= \delta \; \frac{m_i}{N}\frac{N-n_j-m_j}{N \Omega z},
\end{split}
\end{equation}
where the positive constants $\mu$  and $\delta$ quantify the
diffusion ability of the two species and $z$ stands for the number
of first neighbors. To complete the notation
setting, we introduce the $\Omega$-dimensional vectors $ \textbf{n}$
and  $\textbf{m}$ to identify the state of the system. Their $i-th$
components respectively read $n_i$ and $m_i$.

The aforementioned system is  intrinsically stochastic. At time $t$
there exists a finite probability to observe the system in the state
characterized by $\textbf{n}$ and $\textbf{m}$. Let us label
$P(\textbf{n}, \textbf{m},t)$ such a probability. One can then write
down the so called master equation, a differential equation which
governs the dynamical evolution of the quantity $P(\textbf{n},
\textbf{m}, t)$. The master equation for the case at hand takes the
form
\begin{equation}
\begin{split}
\label{spme2}
{\partial \over \partial t}P(\textbf{n},\textbf{m},t) &=
\sum_i^\Omega \Biggl [ (\epsilon_{X_i}^- -1)\;T(n_i+1,m_i|n_i, m_i)
+ (\epsilon_{X_i}^+ -1)\;T(n_i-1 ,m_i, |n_i , m_i)  \\
& +(\epsilon_{X_i}^+\epsilon_{Y,i}^- -1)\;   T(n_i-1,m_i+1 |n_i,
m_i)
+ (\epsilon_{X_i}^-\epsilon_{Y_i}^+ -1)\; T(n_i+1,m_i-1|n_i, m_i)  \\
&+ \sum_{j \in i} \bigl [ (\epsilon_{X_i}^+\epsilon_{X_j}^- -1)\;  T(n_i-1,n_j+1|n_i,n_j)       +\\
&+ (\epsilon_{Y_i}^+\epsilon_{Y_j}^- -1)\;   T(m_i-1,m_j+1|m_i,m_j)     \bigr ] \Biggr ]P(\textbf{n}, \textbf{m}, t),
\end{split}
\end{equation}
where use has been made of the definition of the step operators
\begin{equation}
\label{step}
\epsilon_{X_i}^{\pm} f(\ldots, n_i, \ldots, \textbf{m}) = f(\ldots, n_i \pm 1, \ldots, \textbf{m}), \quad \epsilon_{Y_i}^{\pm} f(\textbf{n}, \ldots, m_i, \ldots) = f(\textbf{n}, \ldots, m_i \pm 1, \ldots),
\end{equation}
and where the second sum in equation  (\ref{spme2}) runs on first
neighbors $j \in i$. Equation (\ref{spme2}) is exact, the underlying dynamics
being a Markov process. The master equation (\ref{spme2}) contains
information on both the ideal mean--field dynamics (formally
recovered in the limit of diverging system size) and the finite $N$
corrections.  To bring into evidence those two components, one can
proceed according to the prescriptions of van Kampen
\cite{vankampen}, and write the normalized concentration relative to
the interacting species as:
\begin{equation}
\label{ansatz}
{n_i \over N} = \phi_i(t) + {\xi_i \over \sqrt{N}}, \quad
{m_i \over N} = \psi_i(t) + {\eta_i \over \sqrt{N}},
\end{equation}
where $\xi_i$ and $\eta_i$ stand for the  stochastic contribution.
The ansatz (\ref{ansatz}) is motivated by the central limit theorem
and holds provided the dynamics evolve far from the absorbing
boundaries (extinction condition). Here $1/\sqrt{N}$  plays the role
of a small parameter and paves the way to a perturbative analysis of
the master equation. At the leading order one recovers the
mean--field equations, while the fluctuations are characterized as
next to leading corrections. The following section is devoted to
discussing the system dynamics, according to its mean--field
approximation.

\section{The mean--field approximation and the Turing instability}
\label{mf} Performing the perturbative  calculation one ends up with
the following system of partial differential equations for the
$\phi_i(t)$ and  $\psi_i(t)$ in cell $i$:

\begin{equation}
\begin{split}
\label{spmf}
\partial_\tau \phi_i =& -b \phi _i-d \phi _i+a \left(1-\phi _i-\psi _i\right)+c \phi _i^2 \psi _i +\mu  \bigl[ \Delta \phi_i + \phi_i \Delta \psi_i  - \psi_i \Delta \phi_i \bigr], \\
\partial_\tau \psi_i =&\;b \phi _i-c \phi _i^2 \psi _i+\delta \bigl[ \Delta \psi_i + \psi_i \Delta \phi_i  - \phi_i \Delta \psi_i \bigr],
\end{split}
\end{equation}
where $\tau=t/(N \Omega)$. In eq. (\ref{spmf}) we  have introduced
the discrete Laplacian operator $\Delta$ acting on a generic
function $f_i$ as
\begin{equation}
\label{discrlap}
\Delta f_i = {2 \over z} \sum_{j\in i} \bigl (f_j - f_i \bigr ).
\end{equation}
The above sum runs on the first neighbors which we assume to total
in $z$. The limit $\Omega \rightarrow \infty$ corresponds to
shrinking the lattice spacing $l$ to zero and so obtaining the
continuum mean-field description. In this limit the system
\eqref{spmf} converges to:
\begin{equation}
\begin{split}
\label{spmf1}
\partial_\tau \phi =& -b \phi-d \phi+a \left(1-\phi-\psi \right)+c \phi^2 \psi _i +\mu  \bigl[ \nabla^2\phi + \phi\nabla^2 \psi - \psi \nabla^2 \phi], \\
\partial_\tau \psi =&\;b \phi-c \phi^2 \psi +\delta \bigl[ \nabla^2\psi + \psi\nabla^2 \phi - \phi \nabla^2 \psi],
\end{split}
\end{equation}
where the population fractions go over the population densities and
the rescaling $\mu \rightarrow \mu l^2$ and $\delta \rightarrow
\delta l^2$ have been performed \footnote{The original parameters
$\mu$ and $\delta$  quantify the ability of the molecules to
exit/enter a given patch of size $l$. When $l$ gets reduced, $\mu$
(resp. $\delta$) should correspondingly increase, so that the
quantity $\mu l^2$ (resp. $\delta l^2$) converges to a finite value
(the physical diffusivity) when the limit $ \Omega \rightarrow
\infty$ is taken. This latter value is in turn the one that enters
the continuous equations (\ref{spmf1}).}. As also remarked in
\cite{LugoMcKane}, cross diffusive terms of the type $\phi\nabla^2
\psi - \psi \nabla^2 \phi$ appear in the mean-field equations, as a
relic of the microscopic rules of interaction. More precisely, they
arise as a direct consequence of the finite carrying capacity
hypothesis. This observation materializes in a crucial difference
with respect to the heuristically proposed reaction diffusion
schemes and points to the need for an extended Turing analysis, i.e.
derive the generalized conditions yielding to organized spatial
structures. Before addressing this specific issue, we start by
discussing the homogeneous fixed point of (\ref{spmf1}). From hereon
we will set $a=d=1$, a choice already made in \cite{boland} for the
original Brusselator model and which will make possible to visualize
our conclusion in the reference plan $(b,c)$ for any fixed ratio of
the diffusivity amount $\delta$ and $\mu$.

\subsection{The homogeneous fixed points}
\label{hsol}

Plugging $a=d=1$ into equations (\ref{spmf1})  and looking for
homogeneous solutions implies:
\begin{equation}
\label{mf2}
\begin{cases}
\dot \phi = (1-\phi-\psi)-(1+b)\phi + c\phi^2\psi, \\
\dot \psi = b\phi - c\phi^2\psi.
\end{cases}
\end{equation}

As an important remark we notice that in the diluted limit $\phi,
\psi <<1$, the system tends to the mean-field of the original
Brusselator model, as e.g. reported in \cite{boland}. For a proper
tuning of the chemical parameters (assigning significantly different
strength) one can prove that, if initialized so to verify the
diluted limit, the system stays diluted all along its subsequent
evolution. As a corollary, the diluted solution is contained in the
formulation of the Brusselator here considered, this latter being
therefore regarded as a sound generalization of the former. Notice
that when operating under diluted conditions,  the cross diffusive
terms appearing in the spatial equations (\ref{spmf1}) can be
neglected and the standard reaction-diffusion scheme recovered.
Again, let us emphasize that it is the finite carrying capacity
assumption that modifies the mean-field description.

System (\ref{mf2}) admits three fixed points, namely:
\begin{equation}
\label{pF}
\begin{cases}
\hat \phi_1 = 0, \\
\hat \psi_1 = 1,
\end{cases}
\begin{cases}
\hat \phi_2 = \frac{c-\sqrt{-8 b c+c^2}}{4 c}, \\
\hat \psi_2 = \frac{1}{2}+\frac{\sqrt{-8 b c+c^2}}{2 c},
\end{cases}
\begin{cases}
\hat \phi_3 = \frac{c+\sqrt{-8 b c+c^2}}{4 c}, \\
\hat \psi_3 = \frac{1}{2}-\frac{\sqrt{-8 b c+c^2}}{2 c}.
\end{cases}
\end{equation}

The first of the above corresponds  to the extinction of the species
$X$: It exists for any choice of the freely changing parameters
$b,c$ and corresponds to a stable attractor for the dynamics. This
solution is not present in the original Brusselator mean--field
equations and its origin can be traced back to the role played
by the limited capacity of the container.

\begin{figure}[!htbp]
\begin{center}
\includegraphics[scale=0.8]{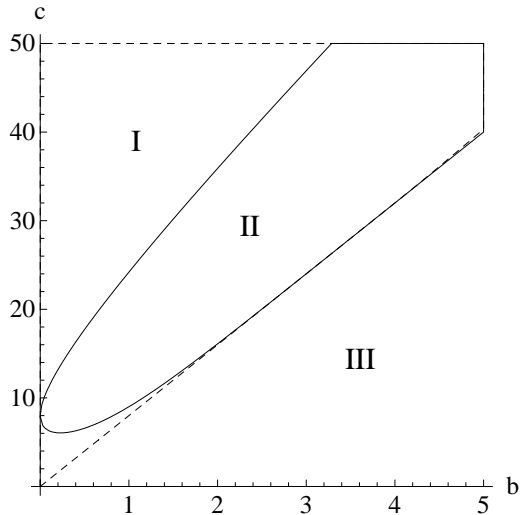}
\caption{The nature of the fixed point $(\hat \phi_3, \hat \psi_3)$
as a function of the parameters $(b,c)$: In region I the fixed point
is a stable node, while in II it is a stable spiral. The dashed line
$c=8b$ delineates region III where the fixed point disappears,
following a saddle node bifurcation. \label{fig_phase_diag}}
\end{center}
\end{figure}

The remaining two fixed points are respectively  a saddle point and
a (non trivial) stable attractor. They both manifest when the
condition $-8b+c>0$ is fulfilled. The saddle point partitions the
available phase space into two domains, each defining the basin of
attraction of the stable points. When $c=8b$ a saddle-node
bifurcation occurs: the fixed points $(\hat \phi_2, \hat \psi_2)$
and $(\hat \phi_3, \hat \psi_3)$ collide to eventually disappear.
The finite carrying capacity that follows the  ``urn representation"
of the Brusselator dynamics destroys the limit cycle solution which
is instead found in its celebrated classical analogue \cite{boland}:
No periodic solutions are found in the mean-field approximation,
unless the diluted limit is  considered. The specific nature of
fixed point $(\hat \phi_3, \hat \psi_3)$ changes as a function of
the parameters  $(b,c)$, as illustrated in figure
\ref{fig_phase_diag}. In region I, it is a stable node, while in II
it is a stable spiral. Region III identifies the parameters values
for which it disappears. We are here  concerned with the stability
of the homogeneous fixed point $(\hat \phi_3, \hat \psi_3)$ to
inhomogeneous perturbation: Can an instability develop and yield to
organized Turing-like spatial patterns?

\subsection{Extending the Turing instability mechanism: the effect of cross-diffusion}
\label{turing1}

To answer the previous question one  has to perform a linear
stability analysis around the selected homogeneous solution $(\hat
\phi_3, \hat \psi_3)$, hereafter termed $(\hat \phi, \hat \psi)$,
accounting for the effect of non homogeneous perturbation. The
formal development is closely inspired by the original Turing
calculation, but now the effects of cross diffusive terms need to be
accommodated for. The technical details of the derivation are
enclosed in the annexed Appendix, where the most general problem is
defined and inspected. We will here take advantage from the
conclusion therein reached and present the results relative to the
Brusselator model. In this specific case $a=d=1$, the dispersion relation
(\ref{pol}) reads in particular:
\begin{equation}
\begin{split}
\lambda(k^2) =& \mathcal{A} + k^2 \mathcal{B} + \frac{1}{16 c} \Biggl[ \mathcal{C}+ \mathcal D k^2 + \mathcal E k^4 \Biggr ]^{\frac{1}{2}},
\end{split}
\end{equation}
where
\begin{equation}
\begin{split}
\mathcal A =& -1+\frac{3 b}{4}-\frac{c}{16}-\frac{1}{16} \sqrt{c (-8 b+c)},\\
\mathcal B =& -\frac{3 \delta }{8}+\frac{\sqrt{c (-8 b+c)} \delta }{8 c}-\frac{\mu }{4}-\frac{\sqrt{c (-8 b+c)} \mu }{4 c},\\
\mathcal C =& 256 c^2+128 b c^2+144 b^2 c^2-32 c^3-32 b c^3+2 c^4+\sqrt{c (-8 b+c)} \left(-32 c^2-24 b c^2+2 c^3\right), \\
\mathcal D =& -64 c^2 \delta +16 b c^2 \delta +8 c^3 \delta +128 c^2 \mu +96 b c^2 \mu -16 c^3 \mu +\sqrt{c (-8 b+c)} \cdot\\
&\cdot\left(-64 c \delta -80 b c \delta +8 c^2 \delta +128 c \mu +32 b c \mu -16 c^2 \mu \right),\\
\mathcal E =& -32 b c \delta ^2+40 c^2 \delta ^2+128 b c \delta  \mu -32 c^2 \delta  \mu -128 b c \mu ^2+32 c^2 \mu ^2+\sqrt{c (-8 b+c)} \left(-24 c \delta ^2-32 c \delta  \mu +32 c \mu ^2\right).
\end{split}
\end{equation}
The perturbation gets amplified if $\lambda(k^2)>0$ over a finite interval of $k$ values. The edges, $k_1$ and $k_2$, of the interval are identified by imposing $\lambda(k^2)=0$ which returns the following results:
\begin{equation}
k_{1,2} = \frac{1}{2 \sqrt{2b \delta  \mu}}\Bigl[2 c \delta (b-1)+8 b^2 (\mu -\delta )-b c \mu  +\sqrt{c (-8 b+c)} (2 \delta -2 b \delta -b \mu ) \pm \bigl(\mathcal F + \sqrt{c (-8 b+c)}\text{  }\mathcal G\bigr)^{1 \over 2}\Bigr]^{1 \over 2},
\end{equation}
where
\begin{equation}
\begin{split}
\mathcal F &= 2 (4 c^2 \delta ^2-8 b c (2+c) \delta ^2+32 b^4 (\delta -\mu )^2-4 b^3 c (8 \delta ^2-2 \delta  \mu +3 \mu ^2)+b^2 c (4 (12+c) \delta ^2+c \mu ^2)), \\
\mathcal G &= 2 (-4 c \delta ^2+8 b c \delta ^2+8 b^3 (2 \delta ^2-\delta  \mu -\mu ^2)+b^2 (-4 (4+c) \delta ^2-16 \delta  \mu +c \mu ^2)).  \\
\end{split}
\end{equation}

\begin{figure}[!htbp]
\begin{center}
\begin{tabular}{cc}
\includegraphics[scale=0.8]{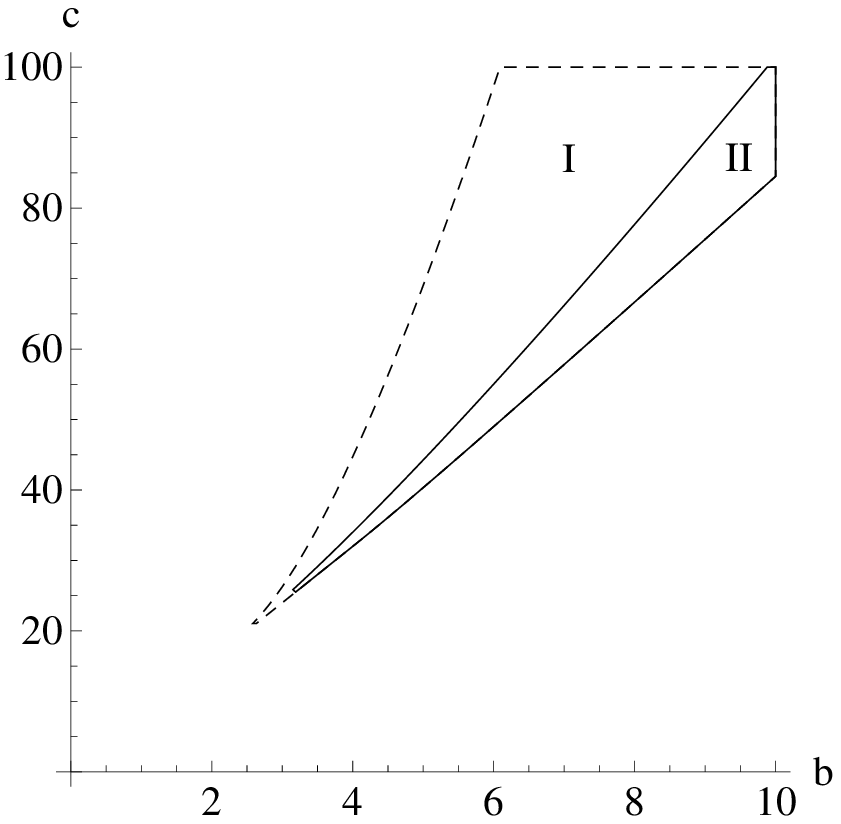} \\
\includegraphics[scale=0.8]{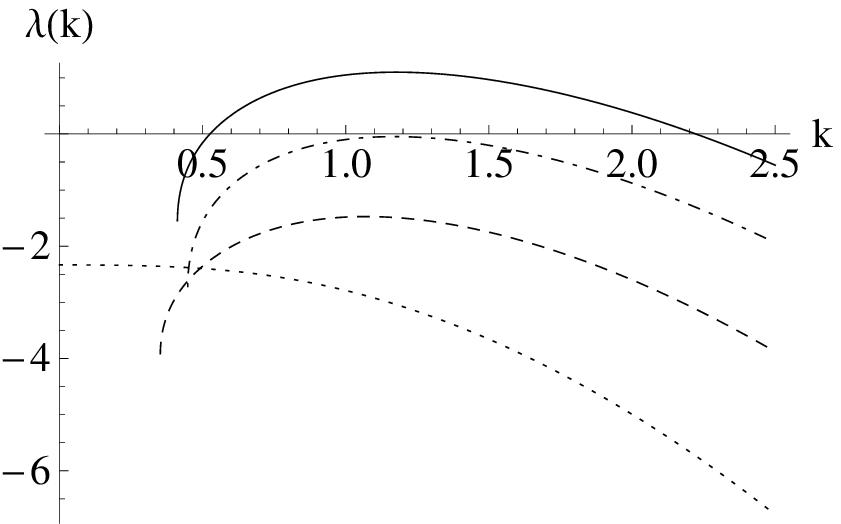} \\
\end{tabular}
\caption{Upper panel: The domain corresponding to the Turing
instability are displayed in the $(b,c)$ plan for $\delta =15$ and
$\mu = 1$. The solid line (zone 2) delineates the region calculated
when accounting for the cross-diffusive contribution, the dashed one
(zone 1) stands for the conventional Turing analysis where diffusion
is modeled via Laplacian operators (i.e. disregarding cross
diffusion). Lower panel: Dispersion relation $\lambda(k)$ vs. $k$.
Here $c=70$ and $b=3,5,6,7$ (starting from the dotted bottom curve).
\label{fig1}}
\end{center}
\end{figure}

By making use of conditions \eqref{turcond} as derived in the
appendix, the predicted  region for the Turing instability is traced
in the $(b,c)$ parameter space for an assigned ratio of mutual
diffusivity, see figure \ref{fig1}. When compared to the
conventional Turing analysis, based on the heuristically
hypothesized reaction diffusion scheme (i.e. removing the cross
terms in eq. (\ref{spmf1})), the region of inhomogeneous instability
shrinks, the condition for spatial self-organization becoming even
more peculiar than so far believed.

Up to now we have carried out the analysis in the  continuous
mean-field scenario neglecting the role of finite size corrections.
Stochastic fluctuations can however amplify due to an inherent
resonant mechanism and consequently give rise to a large scale
spatial order, similar to that predicted within the Turing
interpretative picture \cite{Goldenfeld}. The following section is
entirely devoted to clarifying this important point.

\section{The role of stochastic noise and the power spectrum of fluctuations}
\label{stoch}

At the next to leading order  in the van Kampen perturbative
development one ends up with a Fokker-Planck equation for the
probability distribution of the fluctuations  $\Pi(\boldsymbol{\xi}
, \boldsymbol{\eta},t) = P(\textbf{n},\textbf{m},t)$ where the
vectors $\boldsymbol{\xi}$ and $\boldsymbol{\eta}$ have dimensions
$\Omega$ and respective components $\xi_i$ and $\eta_i$. The
derivation is lengthy but straightforward and the reader can refer
to e.g. \cite{LugoMcKane,DeAnna} for a detailed account on the
mathematical technicalities. We will here report on the outcome of
the calculation in terms of the predicted power spectra of the
fluctuations close to equilibrium. In complete analogy with e.g.
equation (36) and (37) of \cite{LugoMcKane} the latter can be
written as:
\begin{equation*}
P_{k,X} (\omega) = <|\boldsymbol{\xi}_k(\omega)|^2>
\end{equation*}
\begin{equation*}
P_{k,Y} (\omega) = <|\boldsymbol{\eta}_k(\omega)|^2>
\end{equation*}
where the $\omega$ and $k$ stand  respectively for Fourier temporal
and spatial frequencies\footnote{We define the Fourier transform
$f_{ k}$ of a function $f_{\textbf{j}}$ defined on a one dimensional
lattice with spacing $l$ as $f_{k} = l \sum_j \exp(-i k \cdot lj)
f_{\textbf{j}} $, see \cite{LugoMcKane}.} . Exploiting the formal
analogy between the governing Fokker-Planck equation and the
equivalent Langevin equation one eventually obtains (see eqs. (38)
and (39) in \cite{LugoMcKane}):
\begin{eqnarray*}
P_{k,X}(\omega) &=&  \frac{\mathbf{C}_1 + \mathbf{B}_{k,11}\omega^2}{(\omega^2 - \Omega^2_{k,0})^2 + \Gamma^2_{k} \omega^2}, \\
P_{k,Y}(\omega) &=& \frac{\mathbf{C}_2 + \mathbf{B}_{k,22}\omega^2}{(\omega^2 - \Omega^2_{k,0})^2 + \Gamma^2_{k} \omega^2},
\end{eqnarray*}
where:
\begin{eqnarray}
C_{k,X}(\omega) &=& \mathbf{B}_{k,11} M_{k,22}^2 - 2
\mathbf{B}_{k,12} M_{k,12} M_{k,22}
+\mathbf{B}_{k,22} M_{k,12}^2, \\
C_{k,Y}(\omega) &=& \mathbf{B}_{k,22} M_{k,11}^2 - 2
\mathbf{B}_{k,12} M_{k,21} M_{k,11} +\mathbf{B}_{k,11} M_{k,21}^2.
\end{eqnarray}
The $2 \times 2$ matrix $M_k$ of elements $M_{k,ij}$ reads
 \begin{equation}
\mathbf{M}_k =
\left(
\begin{array}{cc}
 -2-b+2 c \hat{\phi } \hat{\psi }+\mu  \left(1-\hat{\psi }\right) \Delta_k & -1+c \hat{\phi }^2+\mu  \hat{\phi } \Delta_k \\
 b-2 c \hat{\phi } \hat{\psi }+\delta  \hat{\psi } \Delta_k & -c \hat{\phi }^2+\delta  \left(1-\hat{\phi }\right) \Delta_k
\end{array}
\right),
\end{equation}
and the elements $\mathbf{B}_{k,ij}$ respectively are:
\begin{align}
\mathbf{B}_{k,11} &=
1+b \hat{\phi }-\hat{\psi }+c \hat{\phi }^2 \hat{\psi }-2 \mu  \hat{\phi } \Delta_k+2 \mu  \hat{\phi }^2 \Delta_k+2 \mu  \hat{\phi } \hat{\psi } \Delta_k, \\
\mathbf{B}_{k,12} &=
-b \hat{\phi }-c \hat{\phi }^2 \hat{\psi }, \\
\mathbf{B}_{k,22} &=
b \hat{\phi }+c \hat{\phi }^2 \hat{\psi }-2 \delta  \hat{\psi } \Delta_k+2 \delta  \hat{\phi } \hat{\psi } \Delta_k+2 \delta  \hat{\psi }^2 \Delta_k,
\end{align}
with the additional  condition
$\mathbf{B}_{k,12}=\mathbf{B}_{k,21}$. Finally, $\Omega^2_{k,0}=det
\mathbf{M}_k$,  $\Gamma_{k}=-tr \mathbf{M}_k$ and the Fourier
transform of the Laplacian $\Delta_k=2  [\cos(kl)-1]$. In the
continuum limit, the cell size goes to zero and $\Delta_k$ scales
as $\Delta_k \sim k^2$ \cite{LugoMcKane}.

\begin{figure}[!htbp]
\begin{center}
\begin{tabular}{cc}
\includegraphics[scale=0.7]{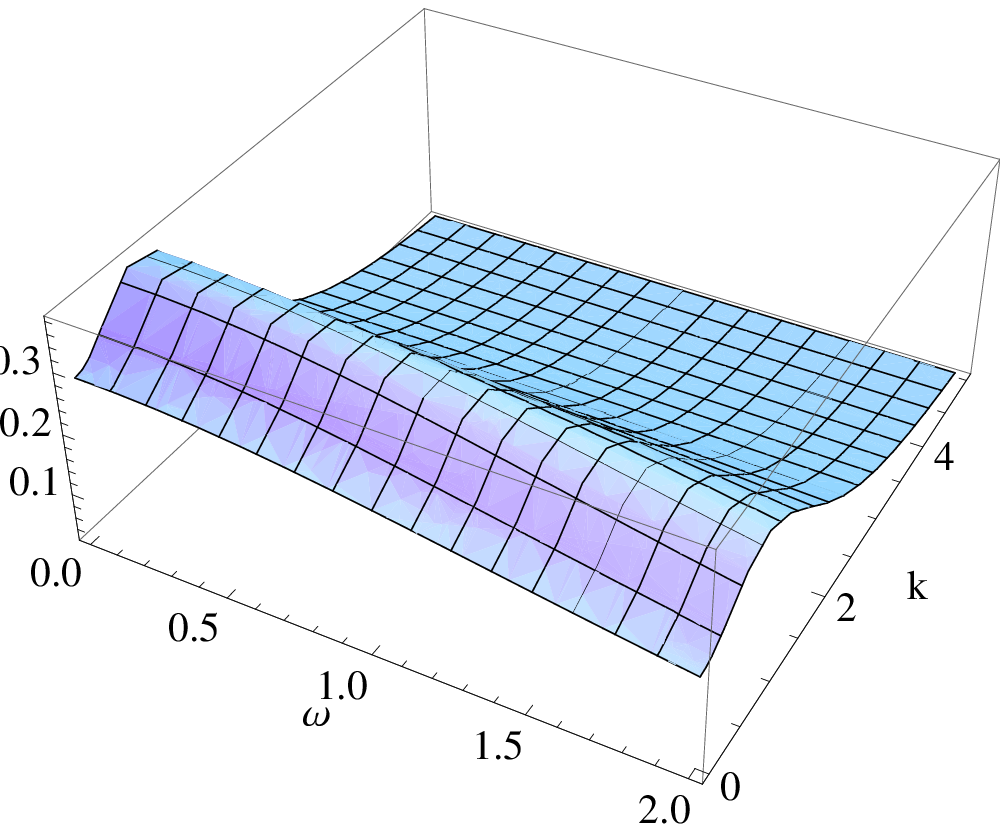} &
\includegraphics[scale=0.7]{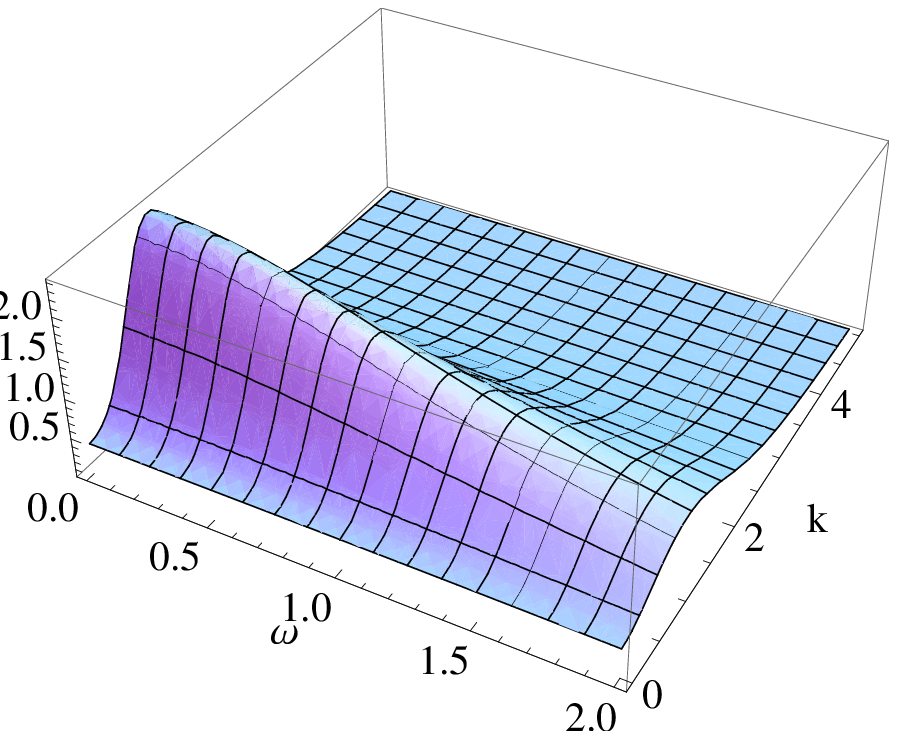}\\
\includegraphics[scale=0.8]{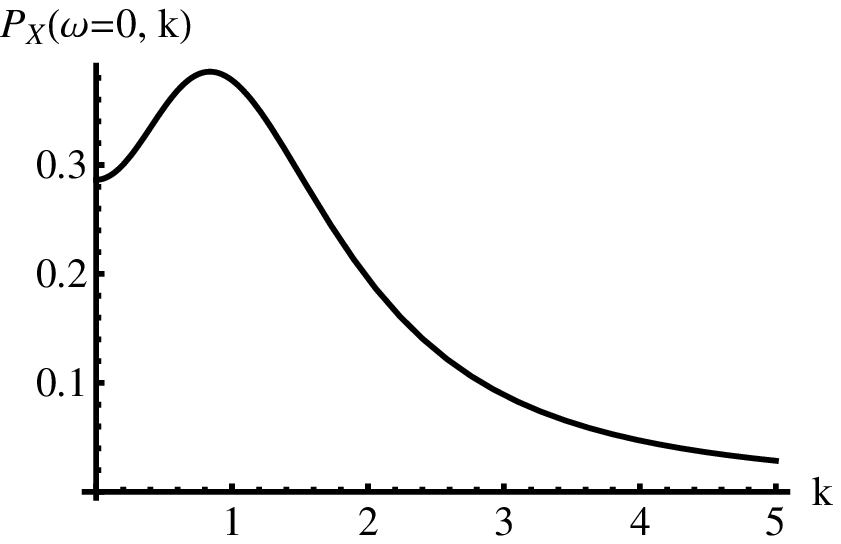} &
\includegraphics[scale=0.8]{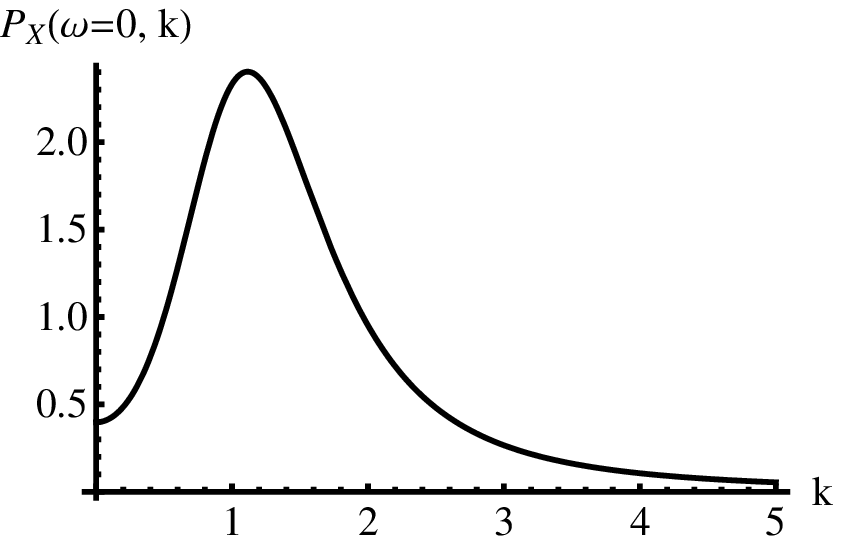}\\
\end{tabular}
\caption{Upper panels: The power  spectra $P_{k,X}(\omega)$ for
$c=70$ and $b=3,6$ (going from left to right). Here $\delta=15$ and
$\mu=1$. The selected values of $b$ and $c$  fall outside the region
of Turing instability, as follows the mean--field linear stability
calculation (see figure \ref{fig1}). Lower panels: The power spectra
$P_{k,X}(0)$ plotted  vs. $k$ so to enable one to appreciating the
range of excited spatial wavelengths. This latter approximately
matches that found in the realm of the Turing mean--field
calculation (see figure \ref{fig1}), implying similar characteristic
of the spatially ordered structures. \label{fig2}}
\end{center}
\end{figure}

\begin{figure}[!htbp]
\begin{center}
\begin{tabular}{cc}
\includegraphics[scale=0.8]{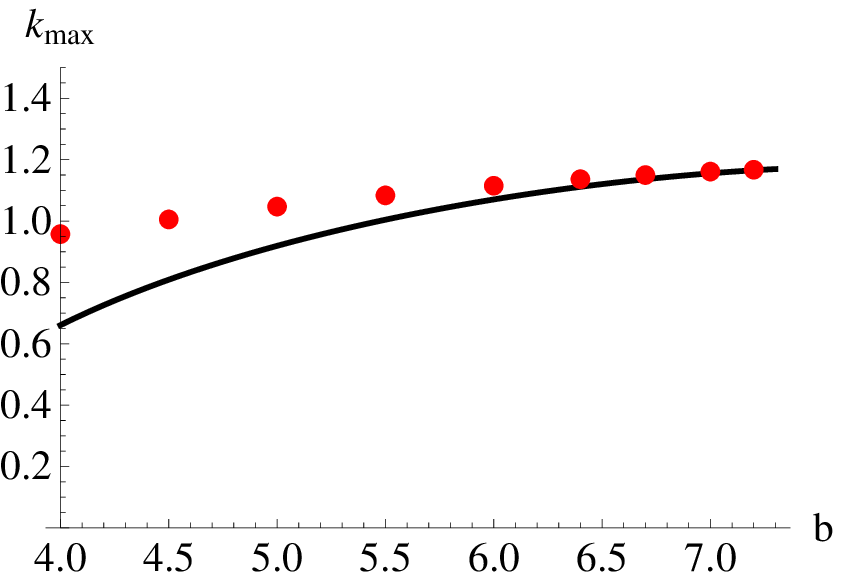} &
\includegraphics[scale=0.8]{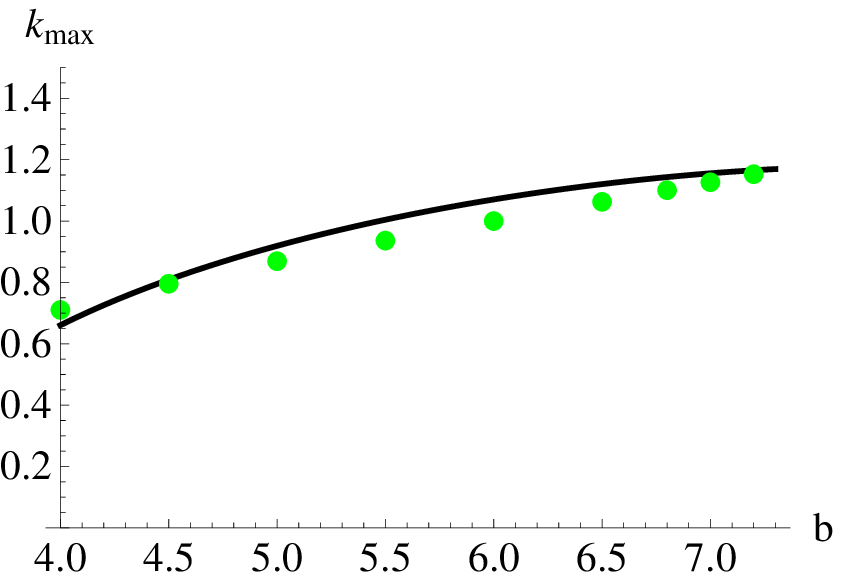} \\
\end{tabular}
\caption{Left panel: Symbols refer  to the values in $k$ that
identify the maximum of the function $P_{k,X}(0)$, here plotted as a
function of $b$. The solid line stands for the position of the
maximum of the mean--field dispersion relation $\lambda(k)$.
Parameters are set as in figure \ref{fig2}. Right panel: as in the
left panel, but now the symbols refer to species $Y$.
\label{fig_kmax}}
\end{center}
\end{figure}

We are now in a position to  represent the power spectrum of the
fluctuations as predicted by the system size expansion. In figure
\ref{fig2} we display a selection of power spectra relative to one
population and  for different values of the parameter $b$, at fixed
$c$ and diffusivity amount (see caption). The snapshots refer to a
region of the parameter space for which we do not expect spatial
order to appear, based on the Turing paradigm. However, and beyond
the simplified mean-field viewpoint, a clear spatial peak is
displayed, which gains in potency as the boundary of the Turing
domain is being approached. Physically, it seems plausible to assume
that the demographic noise can modify the dispersion relation. As a
consequence, the curve $\lambda(k)$, negative defined beyond the
region of Turing instability, can locally cross the zero line,
taking positive values in correspondence of specific $k$. These
latter modes get therefore destabilized, yielding to quasi-Turing
structures. Clearly, the $k$ candidate to drive the stochastic
instabilities are the ones close to $k_{max}$, the wave number that
identifies the position of the maximum of the dispersion relation.
If the above scenario is correct, $k_{max}$ should then be
reasonably similar to the $k$ value that locates the peak position
in the power spectra $P(k, \cdot)(0)$. In figure (\ref{fig_kmax})
such comparison is drawn, for both species and over a window of $b$
values, confirming the adequacy of the proposed interpretation.
Based on the above, we can convincingly argue that the spatial modes
here predicted represent an ideal continuation of the Turing
structures beyond the region of parameters deputed to the
mean--field instability and, for this reason, we suggest to refer to
them as to {\it stochastic Turing patterns}.

\begin{figure}[htbp]
\centering
\vspace*{2.5em}
\includegraphics[scale=0.8]{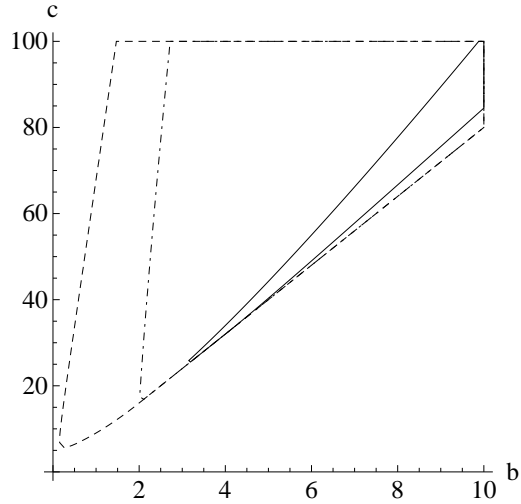}
\caption{\label{fig3}The extended domain  of Turing like instability
predicted by the stochastic based analysis: the dashed line refers
to species $X$, while the dot-dashed to species $Y$. The stochastic
Turing region is confronted to the corresponding mean--field
solution (solid line, also depicted in figure \ref{fig2}).}
\end{figure}

The fact that spatial order  appears for a wider range of the
control parameter is further confirmed by visual inspection of
figure \ref{fig3}, where the region of stochastic induced spatial
organization, delineated from the power spectra calculated above,
is depicted and compared to the corresponding mean--field
prediction. Notice that the domains yielding to stochastic Turing
patterns structures are different depending on the considered
species. This observation marks yet another difference with respect to the
standard mean-field theory. We can in fact expect to observe
spatially ordered structures for just one of the two species, while
the other is homogenously distributed.

\begin{figure}[!htbp]
\begin{center}
\begin{tabular}{cc}
\includegraphics[width=80mm,height=50mm]{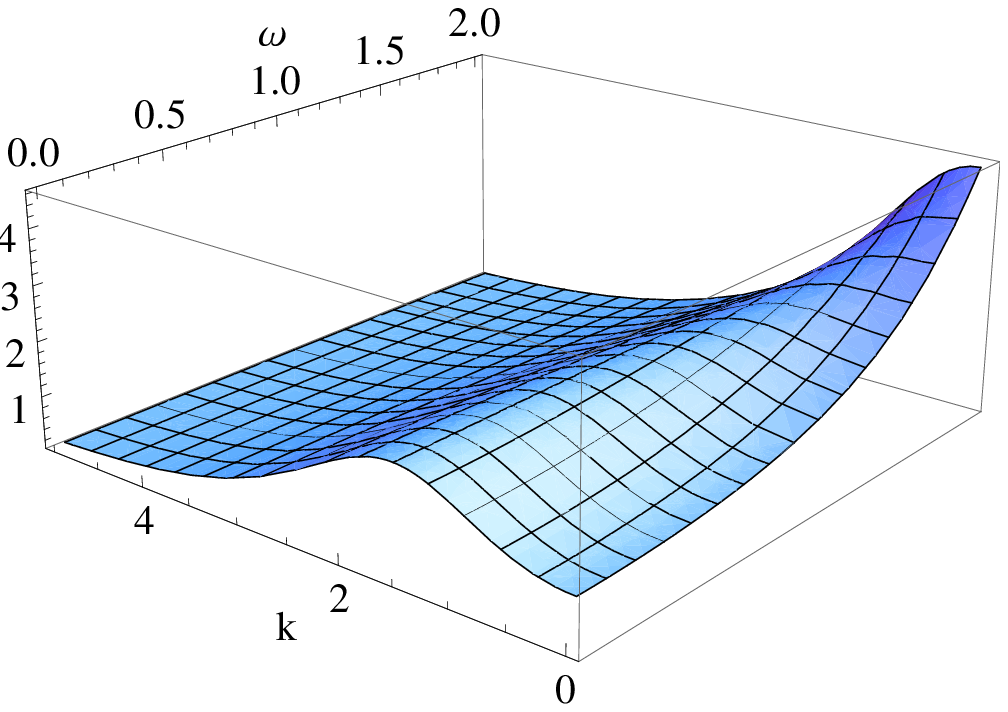} &
\includegraphics[width=80mm,height=50mm]{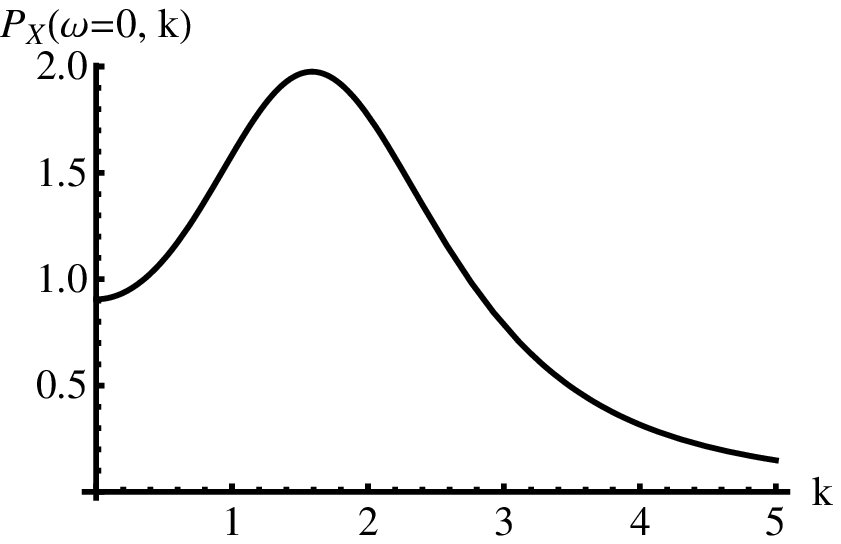}\\
\end{tabular}
\caption{Left panel: The power  spectra $P_{k,X}(\omega)$ for
$a=d=1$, $c=70$ and $b=8$. Spatial and temporal peaks appear now to
be decoupled. Right panel: The corresponding power spectra
$P_{k,X}(0)$ plotted  vs. $k$. A clear peak is displayed.
\label{fig_delta1mu1}}
\end{center}
\end{figure}

As an additional a point,  we consider the case  $\delta=\mu=1$.
This choice cannot yield to the genuine Turing order, a pronounced
difference in the relative diffusivity rates being an unavoidable
pre-requisite. As opposed to the classical picture, stochastic
Turing patterns can still emerge as demonstrated in figure
\ref{fig_delta1mu1}.

From the above expressions for the power spectra, one can also show
that for $\mu=\delta=0$ (the aspatial limit) quasi-periodic time
oscillations manifest in the concentrations. This is an effect of
inner stochastic noise which restores the oscillations destroyed by
the introduction od the carrying capacity. In general terms,
self-oscillatory dynamics could be possibly understood, as resulting
from the discrete nature of the medium, a scenario alternative to
any {\it ad hoc} mean--field interpretation.

 \section{Conclusion}
\label{conc}

Spatially organized  patterns are reported to occur in a large
gallery of widespread applications and are currently interpreted by
resorting to the paradigmatic Turing picture. This vision, though
successful, relies on a mean-field description of the relevant
reaction diffusion schemes, often guessed on purely heuristic basis.
An alternative scenario would require accounting for the intimate
microscopic dynamics and so encapsulating the effect of the small
scale grainess. This latter translates into stochastic fluctuations
that, under specific conditions, can amplify and so give rise to
organized spatio-temporal patterns. In this paper we have considered
a microscopic model of Brusselator, and shown that Turing like
patterns can indeed emerge beyond the parameter region predicted by
the conventional Turing theory and due to the role of finite size
corrections to the mean-field idealized dynamics. This result is
obtained via a system size expansion which enables us to return
closed analytical expressions for the power spectra of fluctuations.
Our results agree with the conclusion reached in \cite{Goldenfeld}
for another model and employing different analytical tools.
Organized patterns can therefore occur more easily than expected, an
observation that can potentially help reconciling theory and
observations. In particular we find stochastic Turing patterns to
emerge for $\delta \sim \mu$, a condition for which Turing order is
prevented to occur.

\begin{acknowledgments}
D.F. wishes to thank Alan J.  McKane for interesting discussions and
for pointing out reference \cite{Goldenfeld}. F.D.P. wishes to thank
Javier Buceta for stimulating discussion. F.D.P thanks financial
support from the ESF Short Visit Grant within the framework of the
ESF activity entitled "Functional Dynamics in Complex Chemical and
Biological Systems".
\end{acknowledgments}

\section{Appendix}

Let us start by the generic set of partial differential equations:

\begin{equation}
\begin{split}
\label{newreacdiff}
\partial_t \phi &= f(\phi, \psi ) + \mu \bigl[\nabla^2\phi + \phi\nabla^2 \psi - \psi \nabla^2 \phi \bigr ], \\
\partial_t \psi &= g(\phi, \psi ) + \delta \bigl[\nabla^2\psi + \psi\nabla^2 \phi - \phi \nabla^2 \psi \bigr ],
\end{split}
\end{equation}
which explicitly allow for cross diffusive  contributions. We choose
to deal with zero flux boundary conditions.

Assume that in absence of diffusion  (homogeneous setting) the model
tends towards a fixed point specified by $(\hat \phi, \hat \psi)$.
In other words, $\hat\phi, \hat\psi$, do not depend over space
variables. Allowing for diffusion, and imposing $\mu \ne \delta$,
can make the system unstable to spatial perturbation. To clarify
this point, let us start by considering the Jacobian matrix for
$\mu=\delta=0$:
\begin{equation}
\mathbf{J} =
\left(
\begin{array}{cc}
f_\phi & f_\psi \\
g_\phi & g_\psi
\end{array}
\right).
\end{equation}
The fixed point $(\hat \phi, \hat \psi)$ is  linearly stable if
$\mathbf{J}$ has positive determinant and negative trace:
\begin{equation}
\label{stabcond}
\det\mathbf J = f_\phi g_\psi - g_\phi f_\psi >0, \quad \text{tr}\mathbf J = f_\phi + g_\psi <0.
\end{equation}
Assuming  \eqref{stabcond} to hold one can  proceed with a
linearization of \eqref{newreacdiff} (with $\mu, \delta$ $\ne 0$)
around $(\hat \phi, \hat \psi)$ and so looking for the sought
condition of instability. Define $\mathbf x = {{\phi - \hat\phi}
\choose {\psi -\hat\psi}}$, then \eqref{newreacdiff} become
\begin{equation}
\label{lineartur}
\partial_t\:\mathbf x = \mathbf J\:\mathbf x + \mathbf{D}\:\nabla^2 \mathbf x, \quad \mathbf D  =
\left(
\begin{array}{cc}
\mu\;(1-\hat\psi) & \mu\;\hat\phi \\
\delta\;\hat\psi & \delta\;(1-\hat\phi)
\end{array}
\right).
\end{equation}
Define the eigenfunctions of the Laplacian operator as:
\begin{equation*}
\label{eigenlap}
\bigl(\nabla^2 + k^2\bigr)\:\mathbf W_k(\textbf{r}) = 0, \quad \textbf{r} \in H,
\end{equation*}
and write the solution to eq.(\ref{lineartur}) in the form:
\begin{equation}
\label{soluz}
\mathbf x(t, \textbf{r}) = \sum_k e^{\lambda t}\:a_k\:\mathbf W_k(\textbf{r}).
\end{equation}
Assume $\mu=1$, or equivalently label with $\delta$  the ratio of the two diffusivities (after proper rescaling of the rate coefficients). Substituting ansatz \eqref{soluz} into eq. \eqref{lineartur}
yields:
\begin{equation*}
e^{\lambda t}\bigl[ \mathbf J-k^2\:\mathbf{D}-\lambda\mathbf{1}\bigr]\mathbf W_k = 0,
\end{equation*}
which implies that the system admits a  solution iff the matrix
$\mathbf J-k^2\:\mathbf{D}-\lambda\mathbf{1}$ is singular, i.e.:

\begin{equation}
\label{det}
\det(\mathbf J-k^2\:\mathbf{D}-\lambda\mathbf{1}) = 0.
\end{equation}

Label the solutions of \eqref{det}  as $\lambda_1(k^2)$ and
$\lambda_2(k^2)$:  They can be interpreted as dispersion relation,
specifying the time scale of departure (or convergence) of the
$k$-th mode towards the deputed fixed point. If at least one of the
two solutions displays a positive real part, the mode is unstable,
and drives the system dynamics towards a non-homogeneous
configuration in response to the initial perturbation. Manipulating
the determinant, \eqref{det} takes the form:

\begin{equation}
\label{pol}
\lambda^2 + \lambda\:q(k^2) + h(k^2) = 0,
\end{equation}
where
\begin{equation*}
\begin{split}
q(k^2) =&\:k^2+ k^2\delta(1 -\hat\phi -\hat\psi) - f_\phi-g_\psi, \\
h(k^2) =&\:k^4 \delta -k^4 \hat\phi\:\delta -k^4 \hat\psi\:\delta - k^2 \delta  f_\phi + k^2 \hat\phi\:\delta  f_\phi+k^2 \hat\psi\:\delta  f_\psi+k^2 \hat\phi g_\phi - \\
&f_\psi\:g_\phi-k^2 g_\psi+k^2 \hat\psi\:g_\psi+f_\phi\:g_\psi.
\end{split}
\end{equation*}
$\lambda_1(k^2)$ and $\lambda_2(k^2)$  are obtained as roots of
\eqref{pol}. The actual dispersion relation, $\lambda(k^2)$ in the
main body of the paper, is the one that displays the largest real part.
We notice that (i) $q(k^2)$ is always positive as
$\delta>0$ by definition; (ii) $(1-\hat\phi-\hat\psi)>0$ as
$\hat{\phi}$ and $\hat{\psi}$ are concentrations; (iii)
$-f_\phi-g_\psi>0$ due to \eqref{stabcond}. Left-hand side of
equation (\eqref{pol}) is hence a parabola with the concavity
pointing upward and the minimum positioned in the half-plane of
negative abscissa. In conclusion a necessary and sufficient
condition for the existence of real roots is  $h<0$ where:
\begin{equation}
\label{h}
h(k^2) = \tilde{a} k^4 + \tilde{b} k^2 + \tilde{c}, \quad
\begin{cases}
\tilde{a} &= \delta(1-\hat\phi-\hat\psi), \\
\tilde{b} &= -\delta f_\phi - g_\psi + \hat\phi(\delta f_\phi + g_\phi) +\hat\psi(\delta f_\psi + g_\psi), \\
\tilde{c} &= \det \mathbf{J},
\end{cases}
\end{equation}
with $\tilde{a}>0, \tilde{c}>0$. The  condition for $h<0$
corresponds to $\tilde{b}<0$ and
$\tilde{b}^2-4\tilde{a}\tilde{c}>0$. Summing up the condition for
the generalized Turing instability reads:
\begin{equation}
\begin{split}
\label{turcond}
&(\delta f_\phi +g_\psi) -\hat\phi\:(\delta f_\phi + g_\phi )- \hat\psi\:(\delta f_\psi + g_\psi ) > 0, \\
&\bigr[ (\delta f_\phi +g_\psi) -\hat\phi\:(\delta f_\phi + g_\phi )- \hat\psi\:(\delta f_\psi + g_\psi ) \bigl]^2 > 4\:\delta (1-\hat\phi - \hat\psi)\det \mathbf J,
\end{split}
\end{equation}
together with \eqref{stabcond}.


\end{document}